\input harvmac

\Title{\vbox{\baselineskip12pt
\hbox{BCCUNY-HEP/02-01} \hbox{hep-th/0201194}}}
{\vbox{\centerline{Velocity-Dependent Forces and an Accelerating Universe}}}
\baselineskip=12pt
\centerline {Ramzi R. Khuri$^{1,2,3}$\footnote{$^a$}{e-mail: khuri@gursey.baruch.cuny.edu.}
and Andriy Pokotilov$^{1,2}$\footnote{$^b$}{e-mail: APokotilov@gc.cuny.edu.
}}
\medskip
\centerline{\sl $^1$Department of Natural Sciences, Baruch College,
CUNY} \centerline{\sl 17 Lexington Avenue, New York, NY 10010}
\medskip
\centerline{\sl $^2$Graduate School and University Center, CUNY}
\centerline{\sl 365 5th Avenue, New York, NY 10036}
\medskip
\centerline{\sl $^3$Center for Advanced Mathematical Sciences}
\centerline{\sl American University of Beirut, Beirut, Lebanon
\footnote{$^{**}$}{Associate member.}}

\bigskip
\centerline{\bf Abstract}
\medskip
\baselineskip = 20pt

In recent work, it was shown that velocity-dependent forces between parallel
fundamental strings moving apart in a $D$-dimensional spacetime
implied an expanding universe in $D-1$-dimensional spacetime. Here we expand
on this work to obtain exact solutions for various string/brane cosmological
toy models.

\Date{January 2002}

\def\D{h}
\def\dx{\dot x}

\def\a{\alpha}

\def\r{\rho}

\def\({\left (}
\def\){\right )}
\def\[{\left [}
\def\]{\right ]}

\lref\prep{M. J. Duff, R. R. Khuri and J. X. Lu, Phys. Rep.
{\bf B259} (1995) 213, hep-th/9412184.}

\lref\dab{A. Dabholkar, G. W. Gibbons, J. A. Harvey and F. Ruiz Ruiz,
Nucl. Phys. {\bf B340} (1990) 33.}

\lref\calk{C. G. Callan and R. R. Khuri, Phys. Lett. {\bf B261}
(1991) 363.}

\lref\astro{A. G. Reiss {\it et al.}, Astron. J. {\bf 116}
(1998) 1009, astro-ph/9805201; S. Perlmutter {\it et al.},
Astrophys. J. {\bf 517} (1999) 565, astro-ph/9812133.}

\lref\carroll{See S. M. Carroll, hep-th/0011110; A. Linde, hep-th/0107176
and references therein.}

\lref\bps{M. K. Prasad and C. M. Sommerfield, Phys. Rev. Lett. {\bf 35}
(1975) 760; E. B. Bogomol'nyi, Sov. J. Nucl. Phys. {\bf 24} (1976) 449.}

\lref\dynam{R. R. Khuri and R. C. Myers, Phys. Rev. {\bf D52} (1995) 6988,
hep-th/9508045.}

\lref\rs{L. J. Randall and R. Sundrum, Phys. Rev. Lett.
{\bf 83} (1999) 3370, hep-th/9905221;  Phys. Rev. Lett. {\bf 83} (1999)
4690, hep-th/9906064.}

\lref\elias{A. Kehagias and E. Kiritsis, JHEP {\bf 9911} (1999) 022.}

\lref\fundst{R. R. Khuri, Phys. Lett. {\bf B353} (2001) 520, hep-th/0109041.}

\lref\liutsey{H. Liu and A. Tseytlin, {\bf JHEP 9801:010} (1998), hep-th/9712063.}

\lref\rust{R. R. Khuri and R. C. Myers, Nucl. Phys. {\bf B466} (1996) 60,
hep-th/9512061.}

\lref\freedman{W. L. Freedman, Phys. Scripta {\bf T85:37-46} (2000), astro-ph/9905222,
and references therein.}

\lref\witten{E. Witten, hep-th/0106109 and references therein.}


\newsec{Introduction}

In a recent paper \fundst, it was shown that the velocity-dependent forces between
parallel fundamental strings in $D$ spacetime dimensions, with certain initial
conditions, lead to an expanding universe in $D-1$ dimensions. The findings were
consistent with recent observations \astro\ of an accelerating universe, and predict
an asymptotially constant late time expansion rate.

We start with the action $S=I_D + S_2$, where
\eqn\sgact{I_D = {1\over 2\kappa^2} \int d^D x \sqrt{-g}
e^{-2\phi} \left(R+ 4(\partial\phi)^2 -{1\over 12} H_3^2\right)}
is the $D$-dimensional string low-energy effective spacetime action and
\eqn\smact{S_2 =-{\mu\over 2} \int
d^2 \zeta \left(\sqrt{-\gamma} \gamma^{\mu\nu} \partial_\mu X^M
\partial_\nu X^N g_{MN} +\epsilon^{\mu\nu} \partial_\mu X^M
\partial_\nu X^N B_{MN}\right)}
is the two-dimesional worldsheet
sigma-model source action. $g_{MN}$, $B_{MN}$ and $\phi$ are the
spacetime sigma-model metric, antisymmetric tensor and dilaton,
respectively, while $\gamma_{\mu\nu}$ is the worldsheet metric.
$H_3 = dB_2$ and $\mu$ is the string tension.
The fundamental string solution to
the combined action, representing stationary macroscopic strings
parallel to the $x^1$ direction, is given by \dab
\eqn\fstring{\eqalign{ds^2 & = \D^{-1} \left(-dt^2+(dx^1)^2\right)
+ \delta_{ij} dx^i dx^j,\cr
e^{-2\phi} & = \D = 1 + {k_n\over r^n},\qquad B_{01} = - h^{-1} \cr}}
where $n=D-4$, $r^2 = x^i x_i$ and
the indices $i$ and $j$ run through the $D-2$-dimensional
space transverse to the string. The constant $k_n = 2\kappa^2 T_1/n\Omega_{n+1}$,
where $T_1=\mu$ is the tension of the string, equal to its mass/length,
and $\Omega_{n+1}$ is the volume of $S^{n+1}$, the $n+1$-dimensional unit sphere.

This solution can be extended to a multi-static string solution
owing to the existence of a zero-force condition. This condition in turn
arises from the cancellation between the attractive gravitational and
dilatational forces of exchange with the repulsive antisymmetric field
exchange, and is based on the existence of supersymmetry and the saturation
of a BPS bound \bps.

It was subsequently shown that, in addition to the zero static force,
the leading order ($O(v^2)$) velocity-dependent forces cancel for moving strings
as well \calk\ (see also \prep).
This result too is associated with the existence of higher supersymmetry \dynam.
Following \dynam, it is straightforward to verify that the four-point amplitude
corresponding to the scattering of two such fundamental string states approaches
zero in the small velocity limit. This is identical to the result found for the
$a=\sqrt{3}$ black holes, which also preserve half of the total spacetime symmetries,
the maximum for such black hole, string or $p$-brane solutions.

The Lagrangian for a test fundamental string moving in
the background of a parallel source string is then given by \refs{\fundst,\calk}
\eqn\lag{{\cal L}=-m \D^{-1} \left(\sqrt{1-\D \dx^2} - 1\right),}
where $m$ is the mass of the string, $\dx ^2 = \dx^i \dx_i$ and
the ``$\cdot$'' represents a time derivative.
It was shown in \fundst\ that the velocity-dependent force following from this
Lagrangian is repulsive whenever the strings are
moving away from each other, and this leads to a further separation of the strings.
Since this type of interaction occurs for any two strings, if we
start with any number of close, parallel strings initially moving
apart in the transverse space, they will continue
to do so indefinitely and will fuel an expanding universe in the
$D-2$-dimensional transverse space and therefore in the
$D-1$-dimensional spacetime orthogonal to the strings. For
example, five-dimensional fundamental strings lead to an expanding
universe in $D=4$ spacetime dimensions.

The toy model presented in \fundst\ consisted of a large number
of fundamental strings initially very close to each other. Each pair of such
strings interacts as above, so that an initial outward propagation of the
strings tends to further push them apart in the transverse
space. In a mean-field approximation, the effective force on each string
was approximated by that of a single, very large source fundamental string whose Noether
charge $k$ is equal to the total charge of all of the strings in
the $D$-dimensional space. The distance $r$ between the test string and the source
string in this model then represents the approximate average position of the
strings, and hence the size of the universe. The
time dependence of $r$ at both early and late times was determined \fundst\
for this expanding model.

In this paper, we expand on the results of \fundst\ to obtain exact solutions
for the radial position as a function of time for the mean-field approximation for
$D=5$ and $D=6$ for the case of zero angular momentum. We also consider a spherically
symmetric toy model and obtain similar results in the very early and late time limits
as well, to give further support to the mean-field approximation results.

\newsec{Generalized $p$-Branes}

Before outlining these solutions, it is interesting to note that the Lagrangian
\lag\ also arises whenever we consider the motion of a maximally supersymmetric
$p$-brane moving in the background of a parallel, identical $p$-brane. For example,
\lag\ is the same Lagrangian one obtains for a test fivebrane moving in the background
of a parallel source fivebrane or for a D0-brane moving in the background of a
source D0-brane (see, e.g., \liutsey). This can be seen immediately either from
supersymmetry, or through dimensional reduction \rust. For the case of the fivebrane,
for example, one replaces the two form $B_{01}$ with a six-form $A_{012345}$ and proceeds
in the same manner to obtain \lag, where now $m$ represents the mass of the test
fivebrane moving in the background of a parallel source fivebrane.

The relevant dimension is the number of transverse dimensions, given by $n=D-p-3$, since the harmonic
function $\D = 1+ k_n/r^n$ depends only on $n$. Here
\eqn\kayen{k_n = {2\kappa_D^2 T_p \over n\Omega_{n+1}},}
where $T_p$, the tension of the $p$-brane, is equal to its mass/$p$-volume \prep.
Compactifying
$q \leq p$ dimensions, we can relate the $D$-dimensional Newton's constant $G_D$ to the
$D-q$-dimensional Newton's constant $G_{D-q}$ via \prep
\eqn\newton{G_D = \kappa_D^2 = \kappa_{D-q}^2 V_q = G_{D-q} V_q,}
where $V_q$ is the compactified $q$-volume. Since $T_p = m/V_p$, it follows that
\eqn\kayentwo{k_n = {2\kappa_{D-q}^2 T_{p-q} \over n\Omega_{n+1}},}
which is just \kayen\ for a $p-q$-brane. In particular, for $q=p-1$, we recover
the string formula. For $q=p$, we obtain the formula for D0-branes. As long as the
longitudinal directions of the branes are held parallel,the dynamics are independent
of $p$, the dimension of the branes, and depend only on the number of transverse directions.

In what follows, we will consider
parallel strings for simplicity, keeping in mind that we could equally
well consider, say, D0-branes in one less dimension.

\newsec{Mean-Field Approximation}

For the case in which we replace the total repulsive force on a single string
by an effective, large string at the origin, it was shown in \fundst\ that
\eqn\rdot{\dot r^2 = {\r (\D \r +2)\over (\D \r +1)^2},}
where $E$ is the constant total energy of the string, and where $\r = E/m$
is the ratio of the energy to the rest energy of the test string. 
We have set the angular momentum $l=0$. 
Following, Chebyshev's Theorem
\footnote{$^1$}{In order to integrate \rdot\ we need to evaluate
integrals of the form
$\int x^m (a + b x^n)^p dx$ (so called binomial
differentials), where $m,n,p$ are any rational numbers
and $a,b$-any constants. Chebyshev proved that integrals of this form can
be expressed through algebraic, logarithmic and inverse circular
functions in only three cases:

1) $p$ is an integer (positive, negative or zero)

2) ${m+1 \over n}$ is an integer

3) ${m+1 \over n}+p$ is an integer.},
only the two cases of $D=5$ ($n=1$) and $D=6$ ($n=2$) can be integrated exactly.

A straightforward integration for $n=1$ yields
\eqn\meanone{\left({\r +3 \over \r +1}\right)\ln \left( \sqrt{r\over a} + \sqrt{{r\over a}+1}\right)
+
\sqrt{r\over a} \sqrt{{r\over a}+1}
= \sqrt{\left(\r+2\right)^3\over \r\left(\r+1\right)^2}
\left({t-t_0\over k_1}\right),}
where $a={\r k_1\over \r +2}$.
For small $r$ (or early time $t$), $r \simeq \left({\r +2 \over \r +1} \right)^2 t^2/k_{1}$,
while for large $r$ (or late time), $r \simeq \sqrt{\r(\r +2)/(\r +1)^2} t$, 
\footnote{$^2$}{The constant energy $E$ does {\it not} include the constant rest
energy $m$. It is straightforward to show that, in the late time limit,
$\gamma = \left( 1- \beta^2 \right)^{-1/2} = \rho + 1$, so that the interaction
energy is ultimately transformed into kinetic energy ($=\left( \gamma - 1 \right) m$).}
both in
agreement with the findings of \fundst. In this three-dimensional transverse space, $r(t)$
represents the mean size of the universe in this toy model.
Restoring factors of the speed of light $c$ in \kayen, it follows that
$k_1 = {2 G_5 M\over L c^2 \Omega_2 } =  {G_4 M \over 2\pi c^2}$,
where $L$ is the length of each string and $M$ is the mass of the source string,
representing the effective total mass of the universe in the mean-field approximation.

In \fundst, it was claimed that $r << k_1$ and $r >> k_1$ corresponded to early and late
times (relative to the current epoch), respectively,
in the expansion of the universe as implied in \meanone. Let us verify this assumption
using estimates of cosmological parameters obtained in \freedman. The current matter density
$\rho$ is very close to the critical density $\rho_0 = 3H_0^2/8\pi G_4$, and the age of the universe
$t_0 = H_0^{-1}$, where $H_0$ is the Hubble's constant at present time. Ignoring numerical factors of
$O(1)$, we take the current size of the universe $r_0 \sim c t_0$ and its mass as $M \sim \rho r_0^3$.
It is then straightforward to show that $k_1$ is at most an order of magnitude less than $r_0$.
It follows that $r << k_1$ corresponds to much earlier times than present
$t << k_1/c$ and $r >> k_1$ corresponds to much later times $t >> k_1/c$. The model for $n=1$ has
the defect that the ratio of relative velocities to relative positions is not immediately a
spatial constant, unless the spatial dimensions of the universe are restricted to two (see next section).

For the more interesting case of $n=2$, a straightforward integration yields
\eqn\meantwo{ \sqrt{r^2 + a} +
\sqrt{a} \left( {\r+2\over \r +1} \right)
\ln \left( {r + \sqrt{r^2 +a} - \sqrt{a} \over r + \sqrt{r^2 +a} + \sqrt{a}}\right)
= \sqrt{{\r(\r +2)\over(\r +1)^2}} (t-t_0),}
where again $a={\r k_2\over \r +2}$ and $t_0$ is a constant.
For small $r$, $r \simeq r_0 \exp{{t \over \sqrt{k_2}}}$, while for large $r$ we again find
$r \simeq \sqrt{\r(\r +2)/(\r +1)^2} t$, both again in agreement with the findings of \fundst.
In this four-dimensional transverse space, the three-dimensional universe may be regarded
as an expanding spherical shell with radius $r(t)$. A subtle point here arises as to
the connection beween $G_5$ and $G_4$. In going from a five-dimensional universe to
a four-dimensional one whose constant time slices consist of the expanding three-sphere,
it follows that $G_5 \sim G_4 r(t)$. So $G_5$ and $G_4$ cannot
both be constant. For constant $G_5$, $k_2 = G_5 M/2\pi^2 c^2$ (from \kayen) is also constant,
but the four-dimensional Newton's law changes with time as the universe expands. The
alternative picture is to demand a constant $G_4$, but then allow for changing $k_2$,
so that the mean-field model in this case should be thought of as an approximation to
a cosmological $p$-brane solution with $k_2$ a function of time. In either case, within the limits
of these toy models, a straightforward calculation shows that $k_2 \sim \bar r^2$, where $\bar r$ is the
current size of the universe. It again follows that the early
($r << \sqrt{k_2}$) and late ($r >> \sqrt{k_2}$) time limits are valid,
as in the $n=1$ case. Especially interesting features of the $n=2$ case, other than allowing for
a spatially constant Hubble's constant, are the inflationary expansion
at early times and the asymptotically constant expansion rate for late times. This latter
feature is generic to these string/brane models, since the velocity-dependent forces
vanish at asymptotically large distances.

\newsec{Spherical Shell Model}

Now consider the following model of a string-seeded universe
for both $n=1$ and $n=2$. $N$ parallel, identical
strings, with $N >>1$, are all located at the same distance $R$ from the center of the transverse
$D-2$ dimensional
space and move with the same, purely radial, velocity $\vec v = \dot R \hat R$ outward from the center. 
We would again like to determine $R(t)$ and to compare our findings with the mean-field
approximation. Before doing so, we note that such a model is consistent with the cosmological
observation of a Hubble's constant. In the $n=1$ ($n=2$) case, the spatial universe consists of
an expanding $2$-sphere ($3$-sphere).
It is straightforward to show that the relative position
is given by $r_{21} = |\vec r_{21}| = |\vec r_2 - \vec r_1|= 2 R \sin \theta/2$, where $\theta$ is the
angle between
the position vectors $\vec r_1$ and $\vec r_2$. Similarly, the relative speed between
two strings $v_{21} = |\vec v_{21}|=|\vec v_2 - \vec v_1|= 2 v \sin \theta/2$, where $\vec v_1$ and $\vec v_2$
are the velocities of the two strings. It follows that $v_{21}/r_{21} = v/R$ is a constant over
each sphere (or for a given time slice), representing the Hubble's constant for this model.

The Hamiltonian for the system of test string moving in the source string
background can be easily obtained from the Lagrangian \lag\ of this system
\footnote{$^3$}{It is convenient to use this velocity dependent form of the Hamiltonian.
In the same way, one can easily obtain the conventional form $H=H(r,p)$ by using the
expression for the momenta $p_i= {m \dx_{i} \over \sqrt{1- h \dx^2}}$. Then
$H={m \over h}\left( \sqrt{1 + h {p^2 \over m^2 }} -1 \right)$ .}

\eqn\ham{H={m \over h} \left( {1 \over \sqrt{1- h \dx^2}}-1 \right),}

For $D=5$ ($D=6$) this model is equivalent to a system of particles on the surface
of a 2-sphere (3-sphere) with two-particle energy of interaction. This interaction
energy is just the difference between the conserved total energy of the 2-particle
system given by \ham\ and the kinetic energy of the test string, since the source
string is  assumed to be stationary in the mean-field approximation. Note also that in the mean-field
approximation (see also \fundst), the energy \ham\ was taken to be a constant, which led
to a solution for the motion of a single test string in the background of a much larger
source string, which approximated the aggregate effect of the velocity-dependent forces
of all the other strings. In the shell model, the interaction energies obtained from \ham\
are not individually constant but must be added into a total energy for the system, which is
then set equal to a constant total energy. The center of mass of the system remains at the
center of the sphere. The interaction energy of one string in the
background of another is then given by
\footnote{$^4$}{Strictly speaking, we should use the more cumbersome
relativistic form of the kinetic energy. However, since most of the subsequent analysis 
involves the early time expansion,
the non-relativistic approximation used for this model is valid. Furthermore,
as we shall see later, the result for late times is not affected by making this
simplification either.}

\eqn\Eint{E_{int12}= {m \over h_{12}} \left( {1 \over
 \sqrt{1- h_{12} {\dot r}_{12}^{2}}} -1 \right) -{m {\dot r}^{2}_{12} \over 2},}
where $h_{12}= 1 + {k_n \over r^{n}_{12}}$ and  $r_{12}$ is the relative position
of the strings.

For $D=5$ we assume for simplicity that the $1^{st}$ string is located at the north
pole of the 2-sphere with radius $R$. Then, by symmetry, the energy of interaction between
the $1^{st}$ string and $dN$ strings which are located inside the belt with
azimuthal angles between $\theta$ and $\theta+d\theta$ is given by

\eqn\dEint{dE_{int12}=dN \left[ {m \over h_{12}} \left( {1 \over
 \sqrt{1- h_{12} {\dot r}_{12}^{2}}} -1 \right) -
{m {\dot r}^{2}_{12} \over 2}\right],}
where $dN={N \over 2} \sin{\theta \over 2} d \theta$, $r_{12}=2 R \sin{\theta
\over 2}$, ${\dot r}_{12}=2 \dot{R} \sin{\theta \over 2}$ and
$h_{12}=1 + {k_1 \over r_{12}}$

In order to obtain the energy of interaction between $1^{st}$ string and all
other strings we need to integrate \dEint\ over $\theta$ from $\theta=0$
to $\theta= \pi$. This can be done explicitly, but leads to a rather complicated,
and not especially illuminating, expression. In order to make a connection with
the early-time mean-field approximation, we first make the
assumption that ${k_1 \over r_{12}}>>1$. This means that the distance between any
two strings $r_{12}<<k_1$
and our model can describe the system during the time when this condition holds,
i.e. early times.

Thus we replace $h_{12} \simeq {k_1 \over r_{12}}$ in \dEint\ . The integration
over $\theta$ is easy to perform, and we obtain for the energy of interaction
between the $1^{st}$ string and all the other strings

\eqn\Ei{E_{1int}=-{4 m N R \over k_1} \left\{ {2 \over 15 a^3}\left[ \sqrt{1-a}~
(3 a^2 + 4 a+ 8)-8 \right] + {a \over 8} + {1\over 3} \right\}}
where
\eqn\a{a={2 {\dot R}^2 k_1 \over R}.} Note that $a$ is restricted to be in the domain
$0 \leq a \leq 1$ for this model to be valid.

By symmetry, the total energy of interaction of $N$ strings is

\eqn\Eti{E_{int}=N E_{1 int}}
and the total conserved energy of the system is

\eqn\Etotal{NE_1 = E =E_{int}+ N {m {\dot R}^2 \over 2},}
where $E_1$ is the total energy of a single string.

First assume $a<<1$ and expand the total interaction energy of the system \Etotal\
in powers of $a$

\eqn\Eexp{E_{int}={3 m R\over 10 k_1} N^2\left[a^2+ O(a^3)\right].}
Thus up to $2^{nd}$ order in $a$, Eq. \Etotal\ takes the form

\eqn\rhoa{\rho={3 N R\over 10 k_1}a^2 + {R\over 4 k_1}a}
where $\rho=E_1/m = E/m N$ is again the ratio of the total energy 
(not including the rest energy) of each string to its rest energy.

Solving the quadratic equation in \rhoa\ for $a$, and using $R<<k_1$, it easily
follows that the linear term in $a$ in \rhoa\ may be dropped. It then follows
that up to an $O(1)$ numerical factor, $a \simeq \sqrt{k_1 \rho \over N R}<<1$.
Since $\rho$ is at least of $O(1)$, it follows that from $a<<1$ that $R >> {k_1 \over N}$.
Dropping the $2^{nd}$ term in the {\it r.h.s.} of \rhoa,
solving \rhoa\ with respect to $\dot{R}$ and integrating we obtain

\eqn\R{R \simeq \left({ \rho\over  k N}\right)^{1/3} t^{4/3}.}

Another domain of interest is when $a \simeq 1$. This corresponds to an even earlier
time, since from \R\ it follows that $a \sim t^{-2/3}$, so that an earlier time corresponds
to larger $a$. For $a\to 1$, the solution $R=R(t)$ is
easily obtained from the definition of $a$ \a\

\eqn\Ra{R \simeq {t^2 \over k_1}}
and is valid for $R \sim k_1/N$, which can be seen
from \Etotal\ with $E_{int}$ given
by \Eti\ and $E_{1 int}$ by \Ei\ with $a \simeq 1$.
Note that this quadratic expansion in time is consistent with
the early-time approximation of the mean-field model.

For large $R$ ($R>>k_1$), we can assume that $h=1+k_1/R \simeq 1$, so that
$\partial H/\partial R =0$. From \dEint , \Etotal\ it then follows that the constant energy
$E_1$ depends only on $\dot R$ and $N$ \footnote{$^5$}{This last statement is obviously
also valid for the correct relativistic expression for the kinetic energy.}
so that
$\dot R$ depends only on $N$ and $\rho$. So the radial velocity $\dot{R}$ is constant.
Thus $R \propto t$ for large $R$, again in agreement with
the mean-field approximation.

As mentioned above, for $D=6$, the transverse motion of parallel strings
is equivalent to the motion of particles with two particle energy
of interaction given by \Eint\ with

\eqn\nexth{h_{12}=1+ {k \over r^{2}_{12}}}
where $r_{12}=2R \sin{\chi \over 2}$,  $\dot{r}_{12}=2 \dot{R} \sin{\chi \over 2}$
and $\chi$ is the azimuthal angle, where we assume that the $1^{st}$ string is located at
the north pole of the 3-sphere.

If we assume (as before) that the $N>>1$ strings are distributed homogeneously on the
surface of the 3-sphere, then the number
of strings located inside the belt with azimuthal angles between $\chi$ and
$\chi + d\chi$ is

\eqn\Nbelt{dN= {2 N \over \pi} \sin^{2}{\chi} d\chi}
and the energy of interaction between the $1^{st}$ string (at the north pole)
and $dN$ strings inside the belt is given by the same expression \dEint\ with
$h_{12}$ given by \nexth\ and $dN$ by \Nbelt. Assuming (as for $D=5$) that
${k_2 \over r_{12}^{2}}>>1$ (i.e. $R<<\sqrt{k_2}$)
and replacing $h_{12} \simeq {k_2 \over r^{2}_{12}}$
we obtain

\eqn\dEnext{d E_{int}={2 N m \over \pi} \sin^{2}{\chi} d\chi
\left[{4 R^2 \sin^{2}{\chi \over 2} \over k_2}\left({1 \over \sqrt{1-{k_2 \dot{R}^2\over
R^2}}} -1 \right) - 2 \dot{R}^2 \sin^{2}{\chi \over 2} \right] }
integrating this expression over $\chi$ from $0$ to $\pi$ we obtain the energy of
interaction between the $1^{st}$ string and all other strings

\eqn\Enint{E_{int}= {2 m N R^2 \over k_2}\left({1 \over \sqrt{1- {
k_2 \dot{R}^2\over R^2}}}-1\right) - m N \dot{R}^2 }
and conservation of energy condition can be written in the same form \Etotal\ as
before with $E_{int}$ given by \Enint\ or

\eqn\Eb{{1 \over \sqrt{1-b}}-1-{b \over 2}=
{\rho k_2 \over 2 N R^2}}
where
\eqn\b{b={k_2 \dot{R}^2\over R^2},}
where $0 < b < 1$.
For $R << \sqrt{k_2/N}$, it follows from \Eb\ that $1-b << 1$.
Alternatively, expanding \Eb\ in powers of $1-b$ and keeping only the first nonvanishing
term we easily obtain that condition $b\simeq 1$ (or $1-b<<1$) leads to
$R<<\sqrt{k_2/N}$. From the definition of $b$ \b\ we see that $R\simeq R_{0}
\exp{{t\over\sqrt{k_2}}}$ in this case, again corresponding to the
exponentially inflationary expansion in the $D=6$ mean-field model.

On the other hand, the condition ${\rho k_2 \over 2 N R^2}<<1$ or
$R>>\sqrt{{k_2 \over N}}$ (again $\rho$ is at least of $O(1)$) 
leads to $b<<1$, which can also be easily seen from \Eb.
As before, expanding \Eb\ in powers of $b$, keeping the lowest nonvanishing
term and then integrating (in order to get  $R=R(t)$) we obtain

\eqn\RR{R \simeq \sqrt{\rho \over 12 N k_2} \, t^2}

For large $R$ (i.e. $R>>\sqrt{k_2}$) we can drop the $1$ in Eq \nexth. In this case we
obviously obtain the same result as for $D=5$ (and generally any $D$): expansion
with constant radial speed, once more in agreement with the mean-field limit.
We emphasize again that the nonrelativistic approximation for the kinetic
energy in both cases does not affect the results for either early or late times.

An interesting possibility in
this case is that the moving strings in $D=6$ lead to an expanding
five-dimensional universe, in which an effective four-dimensional
brane universe resides, following \rs. The asymptotic late time
expansion rate is also
intriguing, and may represent a testable prediction for this type
of model.

One possible advantage to the type of asymptotically flat universe shown in these
models is that, in contrast to a de Sitter universe, S-matrices would be well-defined \witten.
At the same time, these models allow for an accelerating universe without assuming the
existence of a cosmological constant. One can regard the changing acceleration
as corresponding to an effective cosmological ``constant" which varies with time. For example,
the $D=6$ ($n=2$) model which has exponential growth in the early universe, has a constant
effective cosmological constant, $\Lambda \sim k_2^{-1}$ (\carroll)
which, however, is due entirely to the velocity-dependent
forces between the strings/branes. At very late times, the effective cosmological constant
is zero. It is then a straightforward but tedious exercise to determine the exact time
dependence of the effective cosmological constant in these models.

Further investigations of this type of model are clearly
merited. More complicated and far more realistic models, possibly
involving different species of branes could be considered.
Furthermore, it would be interesting to go beyond the analytic, classical
results obtained above, using a possible combination of numerical computations,
quantum string effects and nonequilibrium thermodynamics. Nevertheless, it is
likely that the leading order behaviour of the type of accelerating
string/brane universe considered above is well-described in the classical
approximation. Needless to say, these results await further verification
and a better understanding of the underlying many-body interactions.

{\bf Acknowledgements:} Research supported by PSC-CUNY Grant \# 63497 00 32 and a Eugene
Lang Junior Faculty Research Fellowship.

\listrefs
\end